\documentclass[conference]{IEEEtran}
\usepackage{cite}
\usepackage{amsmath,amssymb,amsfonts}
\usepackage{algorithmic}
\usepackage{graphicx}
\usepackage{textcomp}
\usepackage{xcolor}
\def\BibTeX{{\rm B\kern-.05em{\sc i\kern-.025em b}\kern-.08em
    T\kern-.1667em\lower.7ex\hbox{E}\kern-.125emX}}

\usepackage{enumitem}
\usepackage{listings}
\definecolor{codegreen}{rgb}{0,0.6,0}
\definecolor{codegray}{rgb}{0.5,0.5,0.5}
\definecolor{codepurple}{rgb}{0.58,0,0.82}
\definecolor{backcolour}{rgb}{0.95,0.95,0.92}

\newlist{steps}{enumerate}{3}
\setlist[steps,1]{label=Step \arabic*:, ref=\arabic*, leftmargin=*}
\setlist[steps,2]{label=Step \arabic{stepsi}.\arabic*:, ref=\arabic{stepsi}.\arabic*, leftmargin=14pt}
\setlist[steps,3]{label=Step \arabic{stepsi}.\arabic{stepsii}.\arabic*:, ref=\arabic{stepsi}.\arabic{stepsii}.\arabic*, leftmargin=14pt}

\newtheorem{numclaim}{Claim}
\newtheorem{definition}{Definition}
\newtheorem{protocol}{Protocol}

\begin{document}

\title{Sybil-Resilient Coin Minting}

\author{\IEEEauthorblockN{Ouri Poupko}
\IEEEauthorblockA{\textit{Mathematics and Computer Science} \\
\textit{Weizmann Institute of Science}\\
Rehovot, Israel \\
ouri.poupko@weizmann.ac.il}
\and
\IEEEauthorblockN{Ehud Shapiro}
\IEEEauthorblockA{\textit{Mathematics and Computer Science} \\
\textit{Weizmann Institute of Science}\\
\textit{and Columbia University}\\
ehud.shapiro@weizmann.ac.il}
\and
\IEEEauthorblockN{Nimrod Talmon}
\IEEEauthorblockA{\textit{Industrial Engineering and Management} \\
\textit{Ben-Gurion University}\\
Beer-Sheva, Israel \\
talmonn@bgu.ac.il}
}

\maketitle
\thispagestyle{plain}
\pagestyle{plain}

\begin{abstract}
We describe a distributed coin minting protocol that mints one coin per time unit for each member in a digital community. The protocol assumes that community members use a trust-graph to determine the genuineness of digital identities, and that doing so bounds the number of sybils (fake or duplicate identities) in the community, but does not completely eliminate them. The main goal of the protocol is to be resilient to the sybils that penetrate the community, in the sense that, in the long run, only genuine identities mint coins. The protocol accepts that sybils penetrate the community from time to time (by gaining enough trust within the trust-graph), yet assumes that every sybil is eventually exposed. Since coins minted by a sybil will most probably circulate by the time it is exposed, the protocol puts the responsibility for introducing a sybil onto its trusting neighbours and confiscates subsequent coins minted by them, until the coins minted by that sybil are accounted for. In particular, the protocol confiscates two coins for each coin minted by the sybil:
  one to recover what was wrongly minted and one as a fine for introducing the sybil in the first place.
  We argue that this approach constitutes a mechanism to deter the introduction of sybils into the community and to incentivize sybil hunting (using part of the confiscated money as a reward).
\end{abstract}


\section{Introduction}\label{section: introduction}

The broader scope of this paper is e-Democracy, as it is concerned with social governance methods that are based on the digital realm, and with the quest for distributed methods of governance that do not rely on any central authority. Distributed digital governance of people requires digital identifiers that genuinely identify the participants\cite{shahaf2020genuine}, in order to achieve equality, and in particular one person -- one vote. Digital identifiers may be sybils,\footnote{We use the term as an adjective and hence lowercase.} a term coined by Douceur~\cite{RN320} to represent fake or duplicate identities, which undermine equality. In a central governance, it is the governing body that identifies the members of the governed community. For distributed governance, a decentralized approach will be to construct a trust-graph between the participants, relying on peer-to-peer trust to identify the members of the community. A trust-graph approach requires some agreed upon condition to accept trusted members. Poupko et al.~\cite{RN367} present a trust-graph based method for community growth which is sybil resilient. It does not prevent sybils from penetrating the trust-graph, but rather bounds the amount of those. The protocol presented in this paper builds on top of this method, and therefore assumes a bound on the fraction of sybil penetration.
The definition of a sybil identity is context dependent. We adopt the context of Shahaf et al.~\cite{shahaf2020genuine}, where a genuine identity is the first identity that genuinely represents an individual. A sybil identity is an identity that is not genuine (not genuinely represents, or not the first).

One of the pillars of governance is governing the economy of the community. Cryptocurrencies provide an example of digital currencies that the community can mint and govern without a central authority. We wish to construct a just and egalitarian digital currency~\cite{shahaf2021egalitarian}, in which every member of the community has a right to an equal share of every newly minted coin. However, with the presence of sybils, corrupt people may use them to gain a bigger share of the cake. This is the focus of this paper. We propose an egalitarian currency minting protocol that aims to be sybil resilient. There are two desired outcomes for such a protocol. One is to guarantee that, over time, the amount of coins minted per time unit is equal to the number of genuine identities in the community. The second is to deter sybil creation and incentivize sybil hunting. As we consider it unavoidable that sybils will penetrate the community from time to time, we design the protocol so that once a sybil is exposed, the protocol recovers the amount of coins the sybil minted and burns it (takes it out of circulation). To further deter sybil creation, the protocol collects back double the amount minted by the exposed sybil. It burns the first half and keeps the second half in a public treasury, to be used for rewarding sybil hunters. We show that based on the work of Poupko et al.~\cite{RN367} and assuming reasonable assumptions of the adversarial power of the sybils, our protocol achieves its desired outcomes.

Our notion of sybils and their impact on the community is somewhat different than other works. Common works on sybil attacks~\cite{RN213,RN64,RN345,RN342,RN344,RN359,RN230,RN341} consider the penetration of sybils into existing social networks, like Facebook and Twitter. In this layout it is custom to categorize the identities into two groups of honest identities (genuine) and sybils (fake/duplicate). It is common to regard the honest identities that have social connections with sybils as victims. In the context of e-Democracy we assume a different kind of social network, which is specifically constructed for digital governance. We assume that individuals are actively engaged in keeping the network free of sybils, by signing mutual sureties (creating edges in the trust-graph) to other individuals, only after verifying that their digital identifier is genuine~\cite{shahaf2020genuine}. As such, we adopt the model proposed by Poupko et al.~\cite{RN367}, where the identities are categorized into three groups of honest, corrupt and sybil identities. We consider any identity with an edge to a sybil as corrupt, regardless whether they are indeed the introducers of this sybil, or innocent victims. Based on this model, the protocol in this paper targets the neighbors of a sybil, once exposed, as the source for retrieving back money minted by the sybil. The protocol also assumes that minted money circulates, and therefore it cannot rely on confiscating the minted coins directly, nor any amount found in the digital accounts of the sybil neighbors. It therefore retrieves the collected money by confiscating future minted coins, or more precisely, holding the sybil neighbors from minting until all the sum is paid.

The first assumption of our protocol is therefore that there is a trust-graph among the members of the community, and that it was employed to bound the fraction of sybils in it. The second assumption is that the community has means to expose sybils. We believe this assumption is reasonable, as the community uses the digital identifiers for social interactions and governance. These can vary from centralized, privacy violating, active means, to random encounters between a sybil and a law abiding honest identity. The results of this paper show that as long as the average exposing time of a sybil is shorter than the average life expectancy of a genuine identity, the protocol keeps its goal of fully recovering sybil minted money. We assume that all the identities with edges to a sybil are colluding, and are doing so intentionally. This is not necessarily the case, as honest individuals may endorse sybils by mistake, or because some corrupt individual tricked them. We leave it for the community to handle such cases with other means. For the sake of our protocol, every individual that endorses a sybil is guilty. We therefore conclude, as the protocol divides the fine between all the neighbours of the sybil, that anyone who operates a sybil eventually pays more than she gained.

Once the community exposes a sybil, the protocol assumes that the community has means to verify and approve the act. As a simplified example, the social contract in appendix~\ref{appendix: protocol} collects signatures from the community. If majority of the community signs that the agent is sybil, the protocol marks it as such. We expect real life communities to have their own due process, involving perhaps a  police and a court, for indicting identities as sybils.

\subsection{Related Work}

\paragraph{e-Democracy}
The broader scope of our work is e-Democracy; in this context we mention the use of state-issued digital identities and digital governmental services in Estonia~\cite{RN8}, and in India~\cite{RN327}. These are examples of existing communities (states) with existing governmental structures, moving into the digital realm. On the other direction, exemplifying emerging communities within the digital realm, the Democracy Earth Foundation~\cite{RN194} is a non governmental initiative towards e-Democracy. It is related to Proof of Humanity~\cite{proofofhumanity}, which is quite similar to our approach. They also use a web of trust for identification, though it seems for now that a single endorser is enough for one to be accepted. They use Kleros~\cite{kleros}, a distributed online dispute resolution protocol, to resolve identity disputes. Interestingly, in some cases, when an identity is found to be ‘Duplicate’ or ‘Does not exist’, they remove from the registry all the identities that vouched for it. This is more harsh than the approach presented here, that only penalizes these neighbours. Proof of Humanity also delivers universal basic income to registered users.

\paragraph{sybil attacks}
As digital identities are prone to sybils~\cite{RN320}, we mention survey works on sybil attacks~\cite{RN213,RN230,RN359} and works that propose algorithms for sybils in social networks~\cite{RN64,RN345,RN342,RN344,RN341}. Most works differentiate between sybils and non-sybils. A couple, like Boshmaf et al.~\cite{BOSHMAF2016142} differentiate a third group of victims in between honest identities and sybils. They use learning algorithms to study the social features of the three groups and assign weights on the graph vertices according to these features. They show that this weighted graph enhances the performance of algorithms like sybilRank~\cite{RN341}.

Shahaf et al.~\cite{shahaf2020genuine} introduce the notion of a genuine personal identifier, together with a trust graph of mutual sureties among identities. Based on such a trust graph, Poupko et al.~\cite{RN367} present a method for community growth, by approving new members based on the underlying trust graph's connectivity.

\paragraph{Cryptocurrencies and smart contracts}
We relate to permissionless cryptocurrencies, but mention that these -- e.g., Bitcoin~\cite{RN346} and Ethereum~\cite{RN19} -- do not care for multiple accounts of a single person. Empowered via smart contracts, several cryptocurrencies involve a universal basic income; e.g., GoodDollar~\cite{RN574}, Circles~\cite{RN575}, and more~\cite{RN576}. Smart contracts can also build Decentralized Autonomous Organizations (DAOs)~\cite{ethereum:dao} and several initiatives propose a full framework of DAOs for distributed community governance; e.g., DAOstack~\cite{daostack}, Colony~\cite{colony} and Aragon~\cite{aragon}, which is the same end-goal as ours. All these three offer smart contracts that deploy on Ethereum to govern identities, tokens, organizations (or projects) and decision making processes. They each have their own means to identify members of the community, yet we did not see any reference to the issue of tokens minted by sybils, or punishing sybils through fines. Relying on Ethereum has its caveats. First, with respect of governability, as Ethereum is practically controlled by a small group of miners. Second, with respect to scalability, as Ethereum is a single worldwide ledger.

As another type of smart contracts, Cardelli et al.~\cite{RN371} lay the foundation for a digital social contract, which is ``a voluntary agreement between people that is specified, undertaken, and fulfilled in the digital realm''. Poupko et al.~\cite{RN403} show a distributed, fault tolerant implementation of social contracts. Social contracts add subsidiarity over Ethereum's smart contracts, as each individual is in full possession over the contracts that she runs. This improves on the scalability and governability issues of Ethereum.

Shahaf et al.~\cite{shahaf2021egalitarian} Propose an egalitarian and just cryptocurrency that is minted by the individuals. They assume sybil-free communities and show the conditions under which a currency network of multiple communities can achieve distributive justice asymptotically, that is each identity mints a single coin, equal to all other minted coins, per time unit. Cardelli et al.~\cite{RN371} show examples of how to implement a cryptocurrency through digital social contracts.

\subsection{Paper Structure}

Section~\ref{section: preliminaries} presents the mathematical background for the protocol and section~\ref{section: the protocol} presents the protocol itself. We then analyze it in three constructive steps. Section~\ref{section: static community} starts with a static community, where sybils are exposed gradually, until only genuine identities are left. We show how the fine propagates through the neighbors of the exposed sybils, exactly to the cut in the graph of corrupt identities that separate between sybils and honest identities. Section~\ref{section: spawning sybil} enhances the model by introducing newly generated sybils that immediately replace the exposed ones. We show, that although there are always some not-yet-exposed sybils in the community, in the long run, their share in the minted coins diminishes. Section~\ref{section: probabilistic model} completes the protocol with a more realistic, probabilistic, model where both sybils are exposed and genuine identities cease to exist. The added complexity here is that a genuine, corrupt, identity may cease to exist before paying its fine. We use simulations to show that still, asymptotically, the protocol maintains its goals. Appendix \ref{appendix: protocol} presents a digital social contract implementation of the protocol. Appendix \ref{appendix: simulation} presents the code for the simulations in this paper.

\section{Preliminaries}\label{section: preliminaries}

\paragraph{Communities}
We adopt some definitions of Poupko et al.~\cite{RN367}, with some simplifications. A trust-graph depicts mutual sureties connections between identities. To capture the informal difference between subjective trust (trusting someone to be reliable) and objective trust (trusting the identity to genuinely represent the individual), we use the term \emph{community graph}. This does not affect the formal model, but explains some of the intuition behind the protocol, like why it punishes the middle layer of genuine identities with edges to sybils. The vertices in a community graph are labeled with the true (yet unknown to the protocol) type of the identities they represent. This simplifies the abstract mathematical definition. Implementation wise, the protocol defines corrupt identities (and fine them) from the structure of the graph. The set of identities that the protocol define as corrupt may be a subset of the set of identities that the following definition defines as corrupt.

\begin{definition}[Community Graph]\label{definition: community graph}
A \emph{community graph} is an undirected, finite, labeled graph $G=(V,E)$ with the following characteristics:
    \begin{enumerate}
        \item $G$ is connected.
        \item The label of each vertex is `H', `C' or `S', meaning that the identity represented by this vertex is either honest, corrupt, or sybil.
        \item $\{(v_1,v_2)\in E\ |\ \mathit{label}_{v_1}=\textrm{`H'}, \mathit{label}_{v_2}=\textrm{`S'}\}=\emptyset$; i.e., there are no edges between honest and sybils; put differently, the corrupt identities are those that are genuine but are connected to sybils.
    \end{enumerate}
\end{definition}

A community history captures the dynamics of the graph; and it is modeled as a series (i.e., a sequence) of community graphs.

\begin{definition}[Community History]
A community history is a series of community graphs $G_t=(V_t,E_t)$.
\end{definition}

The following defines the transitions between consecutive steps of a community history. In the simulation, this is just a method that the minting mechanism calls, to simulate changes over time. In the implementation, this is a collection of social interactions between identities, adding and removing mutual sureties to each other.
Mathematically, we use this to abstract-away the changes to the community graph over time.

\begin{definition}[Community transition method]\label{def: comm trans method}
A community transition method is a function $G'=\mathit{Transition}(G)$ that receives a community graph and returns a new community graph.
\end{definition}

\paragraph{Graph connectivity}
To define communities on a community graph, we measure the connectivity of the graph, assuming sybils are less connected than genuine identities. Alvisi et al.~\cite{RN230} show that conductance is an effective graph connectivity measure for bounding the number of sybils. They differentiate between sybils and honest identities, and regard the edges between them as the attack of the adversary. We adopt the model of Poupko et al.~\cite{RN367}, which differentiates between honest, corrupt and sybil identities, and regard the corrupt population as the attack of the adversary. They use vertex expansion, which is similar to conductance, but using the vertex cut, rather than the edge cut in the graph. Assuming that the population of corrupt identities is bounded, i.e., there exists $\gamma$ such that $\frac{|C|}{|V|}\leq \gamma$, vertex expansion bounds the population of sybils.

\begin{definition}[Inner Boundary Vertex Expansion]
    Let $G = (V, E)$ be a graph. Given two subsets $A,B\subseteq V$, the \emph{inner boundary of $A$ w.r.t. $B$} is 
    $$\partial_{v}(A,B):=\#\{x\in A\ |\ \exists y\in B\ s.t.\ (x,y)\in E\}\ .$$
    The \emph{inner boundary vertex expansion} is the minimal normalized inner boundary considering all subsets $A\subset V$ of size at most $\frac{|V|}{2}$ and their complements $A^{c}$:
    $$\Phi(G):=\min_{0<|A|\le\frac{|V|}{2}}{\frac{\partial_{v}(A,A^{c})}{|A|}}\ .$$
\end{definition}

Note that $0\le\Phi(G)<1$, as $\Phi(G)=0$ for a disconnected graph, and $\Phi(G)\le\frac{\delta}{\delta+1}$, for any graph where $\delta$ is the minimum degree of the graph.

\paragraph{sybil penetration ratio}
Poupko et al.~\cite{RN367} guarantee that, given a bound on the population of the corrupt identities $\gamma$ and a bound on the actual measurement of vertex expansion of the graph $\Phi$, the ratio of sybils $\sigma=\frac{|S|}{|V|}$ will retain the bound $\sigma<\frac{\gamma}{\Phi}-\gamma$. We treat the case where corrupt identities and their controlled sybils become majority, as a tilting point for the democratic governance of the community, since once a minority of corrupt identities can be counted as majority (using sybils) they can overcome democratic, majority based, decisions. Shahaf et al.~\cite{ijcai2019-81}. show how to make democratic decisions, while being resilient to a bounded number of sybils, by requiring supermajorities and leaning on the current state of affairs as a status quo. An interesting future work will be to find how their results can help loosen the bounds that this paper assumes.

Additionally, we assume that it is feasible for the community growth mechanism to achieve a bound of $\Phi \ge \frac{2}{3}$ (as shown by the simulations presented by Poupko et al.~\cite{RN367}). The simulations in this paper therefore assume $\gamma=\frac{1}{3}$, which leads to $\sigma \le \frac{1}{6}$, so they use a ratio of 1:2:3 between sybil:corrupt:honest identities. Note that we assume no means to limit the amount of corrupt identities in the community. If the ratio corrupt:honest in a community is more than 2:3 then the community growth mechanism cannot guarantee that corrupt identities and their sybils will remain a minority. As long as the community has faith (or other means of assurance) that the amount of corrupt identities is less than above, we consider the ratio 1:2:3 to be the worst case, as a lower number of corrupt identities will only improve the results given here, and a higher number of sybils will result in a lower value of vertex expansion, and hence will be visible (in particular, the community growth mechanism will not allow it).

\paragraph{Graph traversing}

We assume that all identities in the community graph participate in a joint distributed ledger, with an external trigger that starts a new round periodically. On each round, each identity mints $1$ coin of some currency. When the community exposes sybils, the minting protocol traverses the graph to follow the path of already exposed sybils, until it reaches higher degree neighbors that are still participating in the community (either genuine identities that are corrupt, or sybils that are not yet exposed). For this purpose we define the \emph{conditional boundary} of a vertex, to include all neighbors that meet some condition.

\begin{definition}[Conditional path and boundary]
Given a predicate $\mathit{cond}$ on vertices, a \emph{conditional path} $v \xrightarrow{\mathit{cond}} u$ is a path from $v$ to $u$ where $u$ is the first and only vertex on the path that satisfies $\mathit{cond}$.
The \emph{conditional boundary} $\partial_{\mathit{cond}} v$ of a vertex $v$ is the set of all vertices with a conditional path from $v$ 
$$  \partial_{\mathit{cond}} v=\{u\ |\ \exists\;v \xrightarrow{\mathit{cond}} u\}
\ .$$
If the neighbors of $v$ are changing over time (between rounds), then $\partial_{\mathit{cond}} (v,t)$ is the conditional boundary of $v$ at time $t$.

Similarly, the conditional boundary of a set of vertices is defined as  $\partial_{\mathit{cond}} U:=\cup_{u\in U}\partial_{\mathit{cond}} u$.
\end{definition}

The following sections present and analyze the minting protocol in three steps of simplification.

\section{The Protocol}\label{section: the protocol}

The following is a conceptual description of the protocol. The appendices to this paper describe a detailed implementation:   
  Appendix~\ref{appendix: protocol} presents a possible implementation of the minting protocol as a digital social contract~\cite{RN371};
  while Appendix~\ref{appendix: simulation} presents the simulation code used for the experimental evaluation.
We start by defining the protocol's data structures.

\begin{definition}[Protocol data structures]\label{definition: data structures}
The protocol data structures are as follows.
\begin{itemize}
    \item $x(v)$ is a Boolean depicting whether $v$ is exposed as a sybil or not.
    \item $m(v,t)$ is the amount of money minted by $v$ at time $t$.
    \item $f(v,t)$ is the fine imposed on $v$ for being a neighbour of a sybil at time $t$.
\end{itemize}
\end{definition}

The conceptual description of the protocol is as follows.

\begin{protocol}[The minting protocol]\label{protocol: minting protocol}
The sybil resilient minting protocol is as follows.
    \begin{steps}
        \item On each round $t$, every vertex $v$ where $x(v)=0$ mints 1 coin\label{step: minting step}
        \item For every round $t_2<t$ do the following, as long as $v$ has coins to pay\label{step: fine payment}
        \begin{steps}
            \item If $f(v,t_2)>0$ burn coins up to $\frac{f(v,t_2)}{2}$ and pay a tax up to $\frac{f(v,t_2)}{2}$
        \end{steps}
        \item For every vertex $u$ marked by the community as sybil, the protocol does the following:\label{step: sybil exposure}
        \begin{steps}
            \item $x(u)$ is set to $1$
            \item For every round $t_2$ from 0 to $t-1$ do the following
            \begin{steps}
                \item $\mathit{fine}$ is set as $\mathit{fine}=2 \cdot m(u,t_2) + f(u,t_2)$\label{step: fine calculation}
                \item For each vertex $v \in \partial_{x(v)=0}(u,t_2)$ set\label{step: fine division}
                \begin{displaymath}
                  f(v,t_2)=f(v,t_2)+\frac{\mathit{fine}}{|\partial_{x(v)=0}(u,t_2)|}
                \end{displaymath}
            \end{steps}
        \end{steps}
        \item modify the community graph with user interactions $G_{t+1}=\mathit{Transition}(G_t)$\label{step: community growth}
    \end{steps}
\end{protocol}

Below we give some intuitive explanation for the protocol.
Step~ \ref{step: minting step} is the minting step. Each identity that is not exposed as sybil mints one coin for every round of the protocol. Step \ref{step: fine payment} is the fine payment step. Note that the protocol loops over all past time units and calculates the fine payment for each time unit separately. This is important for correct accountancy, making sure old fines are paid first, and the proofs of the claims in the following sections rely on this order of payment.

Step \ref{step: sybil exposure} handles the exposed sybils. The exposure itself is external to the protocol. It assumes that the community has its means to expose them and notify the protocol of the event (see the contract in Appendix \ref{appendix: protocol} for example of how the community notifies the protocol). Once notified, the protocol first marks the sybil as exposed to prevent future minting. Second, iterating over all previous rounds, the protocol calculates the surrounding, not-yet-exposed boundary of the sybil (separately for each round) and divides the fine (previous fine not yet paid by this sybil, plus double the amount it minted) between the vertices of this boundary. In step \ref{step: community growth} the community modifies the graph between iterations. We abstract this step by calling the $\mathit{Transition}$ method, which only appears in the conceptual description of the protocol. In an actual implementation (see Appendix \ref{appendix: protocol}), steps \ref{step: minting step}-\ref{step: sybil exposure} occur on every start-round event, and step \ref{step: community growth} occurs between rounds, as a collection of events emitted by the individuals, mainly creating edges in the community graph.

\section{A Static Community}\label{section: static community}

We start our analysis of the protocol presented above with a simplified model of a static community -- once started, no further identities leave or join the community. The aim of this model is to show, as identities are exposed and determined to be sybils, that the money they minted is accounted for by the identities on the boundary of the group of sybils, which are corrupt by definition, and eventually fully retrieved as fine. For this model we assume that the community checks exactly one identity in the graph at each round, immediately (deterministically) exposing it if it is a sybil. Indeed, this is quite a powerful assumption; Section~\ref{section: probabilistic model} weakens this assumption to some degree. We also assume that the community prioritises older identities, when picking one for examination, so as to expose older sybils before newly introduced sybils. The third assumption, the assumption that the community remains static, implies formally that in step \ref{step: community growth} the $\mathit{Transition}$ function simply returns the same graph at each round. $\forall t, G_{t+1}=G_t$.

The goal of the protocol is to collect back two coins for every coin minted by a sybil ($2 \cdot m(u,t_2)$ in step \ref{step: fine calculation}), one to balance the amount of coins in circulation (the half that is burned in step \ref{step: fine payment}) and one to punish the endorsers of the sybil (the half that is paid as tax in step \ref{step: fine payment}). It propagates this fine from sybil to sybil ($f(u,t_2)$ in step \ref{step: fine calculation}), until non-sybils pay it. The protocol regards the corrupt identities as responsible for introducing the sybil, this is why step \ref{step: fine division} propagates the fine to the not-yet-exposed boundary of the sybil, until reaching non-sybils with edges to sybils, which are the corrupt identities, by definition.

\subsection{Mathematical Analysis}

We show that the introduction of sybils does not benefit their operators (the corrupt identities) with more money. That is, for every coin minted by a sybil, some neighboring corrupt vertex will pay with a burned coin. In the static community model this result is deterministic and final. At some point in time all sybils are exposed, and in a following point in time all money minted by these sybils is retrieved back. As discussed in the introduction, we acknowledge that corrupt individuals may trick honest individuals to support their sybils, but we leave this for the community to handle. The minting protocol treats every vertex with an edge to a sybil as corrupt and guilty.

\begin{numclaim}[sybil-minted money is eventually burned]\label{claim: fixed community money nullified}
Let $t$ be the round counter in protocol~\ref{protocol: minting protocol}. Let $g$ be the number of genuine identities in the graph $g=|\{v \in V : \mathit{label}_v\neq \textrm{`S'}\}|$. Let $R$ be the amount of coins in circulation and $X$ the amount of tax collected. Then, there exists a round $t_e$ such that $\forall t\ge t_e$ the following holds: $R_t+X_t=t\cdot g$.
\end{numclaim}

\begin{IEEEproof}
We show that the fine induced on the neighbors of a sybil propagates, as some of the neighbors are also sybils, until it reaches a boundary of corrupt identities. We prove by induction on the number of sybils that the corrupt boundary returns exactly twice the amount minted by the sybils. Since the community graph remains the same over time, we treat the cumulative parameters of each vertex as one. That is, $m(v)$ is the total amount minted by $v$, summed over all time units, and $f(v)$ is similarly the total amount of fine induced on $v$.

Assume there is only one sybil $v$ and it has $d$ neighbors. When it is exposed at time $t_l$, it managed to mint exactly $t_l$ coins, and at the point of exposure $f(v)=0$. Therefore the fine imposed on its neighbors is exactly $2\cdot t_l$ (step \ref{step: fine calculation}). Let $t_e$ be a point in time where all fine was paid back. Let $n$ be the number of vertices in the graph and $g=n-1$ the number of genuine vertices. Then by time $t_e$, all vertices except the sybil one minted $t_e\cdot g$ coins. Note that by step \ref{step: fine payment}, for every two coins of fine, one coin is burned and one coin is paid as tax (both coins are considered out of circulation). So the money in circulation is $R_{t_e}=t_e\cdot g+t_l-2\cdot t_l$, that is the money minted by genuine identities, plus the money minted by the sybil, minus the money not accounted due to the fine. Similarly the tax collected is $X_{t_e}=t_l$. Together we get $R_{t_e}+X_{t_e}=t_e\cdot g-t_l+t_l=t_e\cdot g$.

Now suppose by induction that the claim holds for $k$ sybils. Let $S$ be the set of sybils in the graph, $|S|=k+1$, and $C$ be the set of neighbors of sybils that are not sybil by themselves. The number of genuine identities is now $g=n-k-1$. Let $v$ be the last ($k+1$) sybil to be exposed, and $t_l$ be the point in time of its exposure. If $v$ was not a sybil, then by time $t_e$ we would have $R_{t_e}^*+X_{t_e}^*=t_e\cdot(n-k)$. Let $\mathit{fine}^*=\sum_{u\in C}f(u)+f(v)$, be the cumulative registered fine at time $t_l$, just before $v$ is exposed. We have $R_{t_e}^*=R_{t_l}+(t_e-t_l)\cdot(n-k)-\mathit{fine}^*$, and $X_{t_e}^*=X_{t_l}+\frac{\mathit{fine}^*}{2}$. Now let $\mathit{fine}$ be the accumulated fine after $v$ is exposed. We have $\mathit{fine}=\mathit{fine}^*+2\cdot t_l$. We can now calculate:
\begin{align*}
    R_{t_e}&=R_{t_l}+(t_e-t_l)\cdot(n-k-1)-\mathit{fine}\\
    &=R_{t_e}^*-(t_e-t_l)+(\mathit{fine}^*-\mathit{fine})\\
    X_{t_e}&=X_{t_l}+\frac{\mathit{fine}}{2}=X_{t_e}^*-\frac{\mathit{fine}^*-\mathit{fine}}{2}\ .
\end{align*}
Summing together and applying the induction claim we get:
\begin{align*}
    R_{t_e}+X_{t_e}&=t_e\cdot(n-k)-t_e+t_l-\frac{(\mathit{fine}-\mathit{fine}^*)}{2}\\
    &=t_e\cdot(n-k-1)+t_l-t_l=t_e\cdot g\ .
\end{align*}

Since all the sybils are exposed by this point in time, and stop minting, the condition also holds for any $t\ge t_e$. That is, $\forall t\ge t_e,\ R_t+X_t=t\cdot g$, which proves the claim.
\end{IEEEproof}

\subsection{Experimental Analysis}

We build the experimental analysis of the minting protocol according to the three steps of analysis presented in this paper. For the static model, the experimental data only serves as a sanity check to the code of our simulation. We ran a simulation of protocol~\ref{protocol: minting protocol}, with the assumption that sybils are exposed deterministically no more than one at a time, and the assumption that the community graph remains the same from round to round. We ran the simulation with a random community of 120 identities, of which 20 are sybils and 40 are corrupt identities (may be neighbors of sybils in the community graph). The simulation ran for 500 rounds and started by minting 120 coins per round. As expected, after 500 rounds, the total amount of coins (in circulation and as collected tax) was exactly 50,000 coins (500 rounds times 100 genuine, non-sybil identities). That is, any money minted by sybils was eventually recovered and burned. See appendix~\ref{appendix: simulation} for more information on the simulation code.

\section{A Community with Regenerating sybils}\label{section: spawning sybil}

Next we describe a more involved analysis of the protocol, in a more realistic scenario. In particular, in reality communities are not static, but evolve. In the regenerating sybils model we assume that the adversary (assuming the corrupt identities can collude, we regard them all as a single adversary entity) immediately replaces each exposed sybil with a new one. As a result, the community remains saturated with the maximal number of corrupt identities and maximal number of sybil identities (see section \ref{section: preliminaries}), at every step. That is the worst case for the defender (the community that desires to be sybil resilient), under the assumption that genuine identities remain static in the community (an assumption that will be removed in the next section). Starting from this point the protocol can no longer burn all money minted by sybils, as there will always be some sybils not yet exposed. What the regenerating sybils model shows instead is that the money minted by unexposed sybils, together with fine not yet paid, are bounded by a constant. Therefore, as time continues to advance, this constant will become negligible to the amount of money in circulation.

In protocol \ref{protocol: minting protocol} each corrupt identity pays a different fine, according to the structure of the graph. It is possible, theoretically, that some corrupt identities may collect fines faster than they can pay them. Our simulations show that this is not the case. The protocol successfully collects all the fine it induces, after a bounded number of time steps. However, we failed to prove it analytically. If, on the other hand, the protocol divides the fine equally among all the neighbours of all the exposed sybils, then the analytical proof becomes easier. For this purpose we introduce the following modified protocol.

\begin{protocol}[Slightly modified protocol]\label{protocol: modified protocol}
The slightly modified minting protocol is the same as protocol \ref{protocol: minting protocol}, except for the following:
    \begin{steps}[label=Step \ref{step: fine division}:]
        \item Let $W=\partial_{x(w)=0}(u,t_2)\cup\{w \in V_{t2} \ |\  f(w,t_2)>0\}$. For each $w \in W$ set
            \begin{displaymath}
                f(w,t_2)=\frac{\sum_{v \in W}f(v,t_2)+\mathit{fine}}{|W|}
            \end{displaymath}
    \end{steps}
\end{protocol}

The difference between protocol \ref{protocol: minting protocol} and protocol \ref{protocol: modified protocol} is that \emph{all} fine-payers divide the fine equally between them (not only the neighbors of the currently exposed sybil). The simulations show that both versions of the protocol bound the amount of sybil-minted money, as claimed by claim \ref{claim: regenerating claim} (subsection \ref{subsection: regenerating analysis}), so the modified protocol is just for the simplification of the proof of that claim.

\subsection{Mathematical Analysis}\label{subsection: regenerating analysis}

Similarly to the static model, section \ref{section: static community}, we again want to show that the introduction of sybils does not benefit their operators (the corrupts) with more money. Since now the sybil community is regenerating, we can no longer show that the protocol retrieves all their money. We show instead that the money not yet retrieved is bounded by a constant. For a more elegant proof, we generalize the fine in step \ref{step: fine calculation} to be $\mathit{fine}=\alpha \cdot m(u,t_2) + f(u,t_2)$, that is $\alpha>1$ coins (rather than 2) are collected back for each sybil-minted coin. The following claim depicts the relation between this parameter $\alpha$ and the vertex expansion of the graph, and shows that as long as the graph maintains this bound on the vertex expansion, and as long as $\alpha>1$ coins are collected for each sybil-minted coin, then the amount of coins not-yet-collected at any point of time is bounded by the size of the community. It does not grow as $t$ grows.

\begin{numclaim}[sybil-minted money is bounded]\label{claim: regenerating claim}
In protocol \ref{protocol: modified protocol} consider the case where $\alpha=\frac{\phi}{1-\phi}$ and assume $0.5<\phi<1$. Let $\sigma$ be the ratio of sybils in the community. The amount of coins minted by sybils and not yet recovered is bounded by $O(\sigma\cdot n^2)$, where $n=|V_0|$.
\end{numclaim}

\begin{IEEEproof}
Since every vertex is eventually tested for being a sybil, then any sybil~$v$ in the graph at time $t$ is necessarily exposed by time $t+t_l$, where $t_l=|V_0|$ (the size of the graph does not change over time). Also, any sybil neighbor of $v$ at time $t$ is necessarily exposed by time $t+t_l$. It follows that every coin minted by a sybil at time $t$ has turned into fine registered on a corrupt identity by time $t+t_l$ or before, as the fine propagates from sybil to sybil, as they are exposed, until reaching a non-sybil neighbor. Given that the vertex expansion of a community graph is $\phi$, it follows that any set of sybils $S$ has a conditional boundary of non-sybil of size:
$$\frac{|\partial_{\mathit{type}=\textrm{`C'}}S|}{|\partial_{\mathit{type}=\textrm{`C'}}S|+|S|}\ge\phi\ .$$
or:
$$|\partial_{\mathit{type}=\textrm{`C'}}S|(1-\phi)\ge\phi|S|\ .$$
$$|\partial_{\mathit{type}=\textrm{`C'}}S|\ge\frac{\phi}{1-\phi}|S|=\alpha|S|\ .$$
Now, let $S_0$ be the set of all sybils at time $t=0$. The fine for the coins minted by $S_0$ is $\alpha|S_0|$, and by time $0+t_l$, there are at least $\alpha|S_0|$ corrupt identities evenly carrying this fine. As the fine is paid oldest first (by step \ref{step: fine payment}), it follows that the protocol recovers all coins minted at time $t=0$ by sybils (assuming $\alpha\ge 1$) by time $t_l+1$. Iteratively, this holds for any t. Any coin minted at time $t$ is recovered by time $t+t_l+1$. The amount of coins in circulation, minted by sybils and not yet recovered, is at most $\sigma\cdot n\cdot (t_l+1) = O(\sigma\cdot n^2)$.
\end{IEEEproof}

\subsection{Experimental Evaluation}

\begin{figure}[t]
  \centering
  \includegraphics[width=\linewidth]{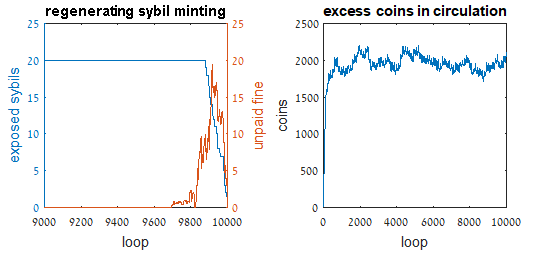}
  \caption{Excess coins in circulation. The diagram on the left shows how many sybils in each round are exposed and how much fine per round is not yet paid. Note that the graph zooms-in on the last 1000 rounds for visibility. The diagram on the right shows the overall amount of sybil coins in circulation in every round.}
\label{figure: regenerating minting}
\end{figure}

We ran the simulation (See appendix \ref{appendix: simulation}) with the assumptions of the regenerating sybils model. The community exposes sybils deterministically, no more than one at a time. Once exposed, the adversary immediately introduces a new sybil into the community. There are two goals for running the simulation. First, to demonstrate the evolution of the fine collecting process, as the community changes over time. Second, to estimate whether the bound proclaimed by claim \ref{claim: regenerating claim} is tight. We ran the simulation with a community of 120 identities, of which 20 are sybils and 40 are corrupt identities. Note that this time the fine is calculated separately for each round, that is for each coin minted by a sybil, the mechanism looks for the not yet exposed neighbors of that sybil, at the time the coin was minted. The payment of the fine is also calculated per round. We ran the simulation for 10,000 rounds to be convinced that the amount of coins in circulation, minted by sybils, is bounded. Note that the simulation divides the fine between the neighbors of the sybil, and as such demonstrates protocol \ref{protocol: minting protocol}, rather then the modified protocol \ref{protocol: modified protocol}. Figure \ref{figure: regenerating minting} shows the results of this simulation. The graph on the left shows on its left axis how many sybils in each round are exposed. It shows that by the time the simulation ended, all sybils older than the last 120 rounds have been exposed, as expected. As a result, coins minted by sybils in the last 120 rounds are still partially in circulation as these sybils are not yet exposed. This is the first source of excess coins in circulation (coins minted by sybils and not yet retrieved). On its right axis the graph shows the fine per round not yet paid. It shows that all fine that derives from coins minted, up to the last about 230 rounds is already paid, while fine that derives from the last 230 rounds is only partially paid. This is the second source of excess coins in circulation. The diagram on the right shows the accumulated amount of excess coins per round. It shows that the amount of excess coins is bounded around roughly 2000, which is slightly less than the bound calculated by claim \ref{claim: regenerating claim}.

\section{A Probabilistic Model}\label{section: probabilistic model}

The last step of the analysis of the minting protocol is the most realistic among the three. First, The assumption that every sybil has a bounded lifespan is too strong. A more relaxed assumption is to assume that every sybil has a probability $p$ to get exposed at every round. Second, the assumption on the immortality of genuine identities is unrealistic, so the probabilistic model assumes that every genuine identity ceases to exist with probability $q$ at every round. We chose the Bernoulli distribution for simplicity. The Gompertz distribution~\cite{wiki:gompertz} better models mortality, but since we don't know what means will the community use to expose sybils, and since we only wish to assess the influence of the ratio between the lifespan of sybils and people on the recovery rate of sybil-minted money, we believe that the simpler model will suffice. These probabilities affect the simulation and not the protocol; as, for the protocol, the death of a genuine identity and the exposure of sybils is an external event.

\subsection{Mathematical Analysis}

We leave the mathematical analysis of the probabilistic model for future work.

\subsection{Experimental Analysis}

\begin{figure}[ht]
  \centering
  \includegraphics[width=\linewidth]{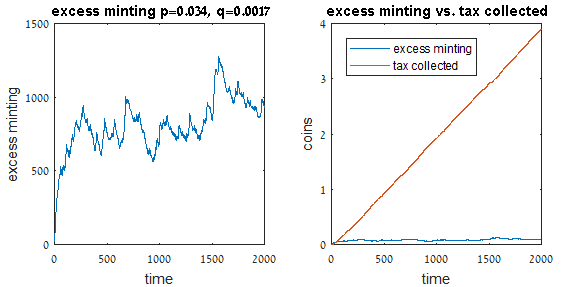}
  \caption{Excess coin minting with sybil exposure probability 0.034 and genuine termination probability 0.0017. On the right is the same graph, compared to the amount of tax collected.}
  \label{figure: p034q0017 excess}
\end{figure}

Again, we construct a community graph with 60 honest vertices, 40 corrupts and 20 sybils, as we consider it to be the extreme case (see section \ref{section: preliminaries}). The simulation maintains this ratio all the time. Whenever it removes a vertex from the graph, it introduces a new vertex of the same type (but with different neighbors). Appendix \ref{appendix: simulation} shows the simulation code. Once genuine identities become mortal, it is no longer guaranteed that the protocol will retrieve double the amount of coins minted by sybils, as every time a corrupt identity ceases to exist, any debt it did not yet pay is lost. We roughly estimate that the average time span of a sybil in a real world community will be no more than several months (maybe a few years), as we assume the community has means to expose sybils as they socially interact. On the other hand, we expect the average time span of a genuine identity to be in the order of tenths of years (as the life expectancy of the individual it represents). We therefore estimate that the ratio $q\approx p/20$ is reasonable and chose $p=0.034$ and $q=0.0017$, which gives a half life of about 20 loops for a sybil (before being exposed) and a half life of 400 loops for a genuine identity.

\begin{figure}[ht]
  \centering
  \includegraphics[width=0.5\linewidth]{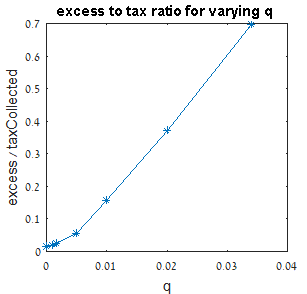}
  \caption{The ratio of excess minting to tax collected for different $q$ values.}
  \label{figure: p034 varq}
\end{figure}

Figure \ref{figure: p034q0017 excess} shows the excess minting for this run. It shows that the amount of excess money grows slowly and steadily, as corrupt identities ceases to exist and any fine they did not yet pay is lost. However, the graph on the right shows that the excess money minted is negligible compared to the money collected as tax. We conclude that the protocol still retrieves and burns all the money minted by sybils, and also collects almost all the desired tax.

The last run of the simulation tested several $q$ values in the range $0\le q\le p$. Figure \ref{figure: p034 varq} shows the ratio between the excess minting to the amount of tax collected. It shows that as long as the probability that a genuine identity will cease to exist is not higher than the probability to expose a sybil, the community can still burn all money minted by sybils.

\section{Outlook}

We presented a community coin minting protocol that is resilient to a bounded fraction of sybils entering the community. The protocol assumes a trust-graph (in a form that we define as a community graph) between identities that bounds the number of sybils to begin with. It further assumes that the community has means to expose and indict sybils from time to time. It then uses the structure of the community graph to collect back sybil-minted money from the identities that endorsed the sybils. We showed the conditions under which the protocol succeeds to fully retrieve sybil-minted money, and even collect additional sum as punishment against the introduction of sybils and as a reward for sybil hunting. A future research direction would be to study our situation as a Stackelberg game between an attacker (say, a wealthy oligarch) and a defender (possibly, the community as a whole).

\bibliographystyle{IEEEtran}
\bibliography{Democracy}
\pagebreak

\appendices

\section{A Social Contract for the Minting Protocol}\label{appendix: protocol}

Following is an implementation in python for a digital social contract that implements the sybil resilient minting protocol as discussed in this paper.

\begin{lstlisting}[language=Python,basicstyle=\small]
# The Currency contract maintains a list of
# accounts for the Community contract
# - external trigger starts minting rounds
# - on exposing sybil the contract punishes
#   its neighbours

class Currency:
 def __init__(self):
  # Storage is an external interface
  self.accounts = Storage('accounts')
  # params is an external interface
  if params.get('community') is None:
   params.update({'community': None,
    'tax_collected': 0, 'timestamps': []})

 # supply an external community contract
 def initialize(self, community):
  params.update({'community': community})

 # the community adds members manually
 def add_member(self, member):
  community = params.get('community')
  if community is None:
   # contract not initialized yet
   return
  if member in self.accounts:
   # skip if member already exists
   return
  if member in community:
   # initialize member's record
   record = {'exposed': False,
    'balance': 0, 'fine': []}
   self.accounts[member] = record

 # assume external mechanism for agent id
 def check_approvals(self, approvals):
  approval_count = 0
  for approval in approvals:
   if approval in self.accounts:
    approval_count += 1
  # tx accepted when majority approves
  if 2*approval_count > len(self.accounts):
   return True
  return False

 # members report a sybil, pending majority
 def report_sybil(self, member, approvals):
  community = params.get('community')
  if community is None:
   # contract not initialized yet
   return
  if self.check_approvals(approvals):
   self.accounts.update(member,
    {'exposed': True})
   # calculate fine
   fine_vec = self.accounts[member]['fine']
   timestamps = params.get('timestamps')
   for index, value in enumerate(fine_vec):
    # fine twice the amount minted plus
    # unpaid fine
    fine = 2 + value
    # find non exposed neighbours
    candidates = [member]
    non_exposed = []
    exposed = []
    while candidates:
     new_candidates = []
     for candidate in candidates:
      if candidate in exposed or\
         candidate in non_exposed:
       continue
      account = self.accounts[candidate]
      if account['exposed']:
       new_candidates.append(
        community.get_neighbors(
         candidate, timestamps[index]))
       exposed.append(candidate)
      else:
       non_exposed.append(candidate)
     candidates = new_candidates
    # divide fine between non_exposed
    # neighbors
    if non_exposed:
     fine = fine/len(non_exposed)
    for neighbor in non_exposed:
     account = self.accounts[neighbor]
     vector = account['fine']
     vector[index] += fine
     self.accounts.update(neighbor,
      'fine', vector)

 # report ceased members, pending majority
 def report_dead(self, member, approvals):
  if params.get('community') is None:
   # contract not initialized yet
   return
  if self.check_approvals(approvals):
   del self.accounts[member]

 # an external trigger triggers minting
 def start_round(self, timestamp):
  if params.get('community') is None:
   # contract not initialized yet
   return
  params.append('timestamps', timestamp)
  for member in self.accounts:
   if self.accounts[member]['exposed']:
    # exposed sybils don't mint
    continue
   # mint at most one coin per round
   minted = 1
   fine_vec = self.accounts[member]['fine']
   # check for induced fine
   for index, value in enumerate(fine_vec):
    payment = min(minted, value)
    params.update_inc('tax_collected',
     payment)
    fine_vec[index] -= payment
    minted -= payment
    if minted == 0:
     break
   # add what is left after paying the fine
   self.accounts.update_inc(member,
    'balance', minted)
   self.accounts.update(member, 'fine',
    fine_vec)
\end{lstlisting}

\pagebreak

\section{The Simulation Code}\label{appendix: simulation}

The following is the Matlab code used for the simulations of protocol \ref{protocol: minting protocol}. This is the actual code for section \ref{section: probabilistic model}. The code for sections \ref{section: spawning sybil} and \ref{section: static community} is similar, with some simplifications. Lines 12-25 initialize data structures. Specifically, line 15 calls the graph generation function (listed below). The protocol runs in a loop for the requested number of rounds (line 27). The first part (lines 28-51) does the minting, by first allocating one coin for every agent (line 33), and then deducting from that coin any past debt (lines 36-48). The remainder remains in the hands of the agent (line 50). The second part starts by tossing a coin for each agent, according to the given probabilities, to expose sybils and terminate deceased genuine identities (lines 53-59). Then, looping over the exposed or deceased agents (line 61), the protocol takes them out of the graph (line 69, 108) and (looping over past time - line 73) propagates through the neighbours of the exposed sybils (line 80-95). It then divides the sybil's penalty between the neighbours that are not themselves exposed as sybils (lines 96-104). At the last step the protocol calls the graph generation function again to fill the graph with new identities, replacing the deceased and exposed ones (line 122).

\begin{lstlisting}[language=Matlab,basicstyle=\small]
function [mintHist,treasury,dead,inDebt] =
  Minting(honest,corrupt,sybil,degree,
          rounds,expProb,deathProb)
%MINTING Simulates sybil resilient minting
%  money with an HCS graph
%   A - adjacency matrix of an HCS graph
%   degree - maximum degree of the graph
%   corrupt - number of corrupt identities
%   sybil - number of sybil identities
%   rounds - number of rounds to run

  % initialize the history matrix
  nodes = honest+corrupt+sybil;
  A=zeros(nodes,nodes,rounds);
  A(:,:,1)=GrowRandomGraph(A(:,:,1),degree,
                           corrupt,sybil);
  mintHist = zeros(rounds,nodes);
  accountedHistory = zeros(rounds,nodes);
  inDebt = zeros(rounds,nodes);
  dead = zeros(rounds,nodes);
  treasury = 0;
  lastCorrupt=nodes-sybil;
  sybilVec = [zeros(1,honest+corrupt),
              ones(1,sybil)];
  shifter = repmat(1:nodes,rounds,1);

  for round=1:rounds
    % tag debts
    inDebtMint = double(any(
                  inDebt.*(~dead)));
    % mint money
    mintHist(round,inDebtMint==0) = 1;
    accountedHistory(round,:) = 1;
    % half the debt acts as punishment and
    % half annihilates excessive money
    for secRound = 1:round
      if ~any(inDebtMint)
        break;
      end
      payments = min([inDebt(secRound,:);
                      inDebtMint;
                      ~dead(secRound,:)]);
      inDebt(secRound,:) = 
        inDebt(secRound,:) - payments;
      treasury = treasury +
                 sum(payments)/2;
      inDebtMint = inDebtMint - payments;
    end
    % the agent keeps what wasn't paid
    mintHist(round,:) = 
      mintHist(round,:) + inDebtMint;
        
    % remove exposed sybils and deceased
    % identities according to probabilities
    probs=rand(1,nodes);
    probs(1:lastCorrupt) =
      (probs(1:lastCorrupt)<deathProb);
    probs(lastCorrupt+1:end) =
      (probs(lastCorrupt+1:end)<expProb);

    for v = find(probs)
      % in case of an exposed sybil
      if(v>lastCorrupt)
        % calculated debt and mark exposed
        debt = accountedHistory(:,v)*2 +
               inDebt(:,v);
        accountedHistory(1:round,v) = 0;
        inDebt(1:round,v) = 0;
        dead(1:round,v)=1;
                
        indexes = find(debt);
        % for each round with a debt
        for index=1:length(indexes)
          secRound=indexes(index);
          % find active neighbours
          neighbours =
            (A(v,:,secRound)==1);
          visited=neighbours;
          visited(v)=1;
          while ~isempty(find(neighbours &
              dead(secRound,:) &
              sybilVec, 1))
            secondNeighbours = 
              any([A(neighbours &
                     dead(secRound,:) &
                     sybilVec,:,secRound);
                   zeros(1,nodes)]);
            secondNeighbours =
              secondNeighbours & ~visited;
            neighbours = (neighbours &
                (~dead(secRound,:) |
                ~sybilVec)) |
                secondNeighbours;
            visited = visited | neighbours;
          end
          % fix their debt
          if(sum(neighbours &
                 ~dead(secRound,:))>0)
            neighbours = (neighbours &
                ~dead(secRound,:));
          end
          inDebt(secRound,neighbours) =
            inDebt(secRound,neighbours) +
            debt(secRound)/sum(neighbours);
        end
      else
        % a deceased genuine identity
        dead(1:round,v)=1;
      end

      % prepare next round
      if round<rounds
        A(:,:,round+1)=A(:,:,round);
        A(v,:,round+1) = 0;
        A(:,v,round+1) = 0;
      end
    end

    % reconnect graph with new identities
    if round<rounds
      A(:,:,round+1) =
        GrowRandomGraph(A(:,:,round+1),
            degree,corrupt,sybil);
    end
  end
end
\end{lstlisting}

We list below The graph generator function. It is a random graph generator that picks at random two vertices which degree is less than the target degree (lines 18-27), and connects them (lines 48-52) as long as these are not an honest and a sybil vertex (lines 28-35). It then continues in a loop until all vertices have the required degree (line 17). The generated graph has maximal degree $d$ and minimal degree $d-1$. Such a graph is slightly simpler to construct compared to a $d$-regular graph, and it seems adequate enough for these simulations. The method also cleans the graph at every round by removing redundant edges between two vertices with degree $d$ (lines 53-55). Our simulations use this mechanism to construct both the initial graph $G_0$, as well as the next iteration of the graph (The $\mathit{Transition}$ method), after each loop (after the exposed sybils and terminated genuine identities are removed from the graph).

\begin{lstlisting}[language=Matlab,basicstyle=\small]
function A = GrowRandomGraph(A,degree,
                             corrupt,sybil)
%GROWRANDOMGRAPH Simulates a trust graph of
%  sybil, corrupt and honest identities
%   A - adjacency matrix of an HCS graph
%   degree - maximum degree of the graph
%   corrupt - number of corrupt identities
%   sybil - number of sybil identities

  % initialize an adjacency matrix
  nodes = size(A,1);
  failedAttempts=0;
  lastHonest=nodes-sybil-corrupt;
  lastCorrupt=nodes-sybil;

  % loop while minimal degree < 'degree'-1
  while min(sum(A)) < degree-1
    % pick a random node with few edges
    desolatedNodes = find(sum(A)<degree-1);
    v = desolatedNodes(randi(length(
          desolatedNodes)));
        
    % pick nodes with room for another edge
    goodNodes = (sum(A)<degree);
    % that are not connected to v
    goodNodes(v)=0;
    goodNodes(logical(A(v,:)))=0;
    % if sybil, don't connect to honest
    if v > lastCorrupt
      goodNodes(1:lastHonest)=0;
    end
    % if honest, don't connect to sybil
    if v <= lastHonest
      goodNodes((lastCorrupt+1):end)=0;
    end
    goodNodes = find(goodNodes);
        
    % if no good nodes, report an error
    if(isempty(goodNodes))
      disp('Failed to find a partner');
      failedAttempts = failedAttempts+1;
      if(failedAttempts>1000)
        break;
      end
      continue;
    end
        
    % pick one at random and connect
    u = goodNodes(randi(length(
          goodNodes)));
    A(u,v)=1;
    A(v,u)=1;
    B = repmat(sum(A),nodes,1) +
        repmat(sum(A,2),1,nodes);
    A(B==degree*2) = 0;
  end
end
\end{lstlisting}

\end{document}